\documentclass[conference]{llncs}
\usepackage[latin1]{inputenc}
\usepackage{amsmath}
\usepackage{amsfonts}
\usepackage{amssymb}
\usepackage{fancyvrb}
\usepackage{graphicx}
\usepackage{listings}
\usepackage{epsfig}
\usepackage[nooneline]{subfigure}
\usepackage{hyperref}
\usepackage{wrapfig}
\usepackage{tikz}
\usetikzlibrary{fadings}
\usepgflibrary{shapes.misc}
\usetikzlibrary{automata}
\usetikzlibrary{positioning}
\usetikzlibrary{trees}

\usepackage{paralist}
\newcommand{\Dynamic}{{\textcolor{blue}{\bf \tt Dynamic}}}
\newcommand{\Assignment}{{\textcolor{blue}{\bf \tt  Assignment}}}
\newcommand{\ParallelAssignment}{{\textcolor{blue}{\bf \tt  ParallelAssignment}}}

\newcommand{\Composition}{{\textcolor{blue}{\bf \tt  Composition}}}
\newcommand{\System}{{\textcolor{blue}{\bf \tt  System}}}
\newcommand{\Controller}{{\textcolor{blue}{\bf \tt  Controller}}}
\newcommand{\Plant}{{\textcolor{blue}{\bf \tt  Plant}}}
\newcommand{\Continuous}{{\textcolor{blue}{\bf \tt  Continuous}}}
\newcommand{\Discrete}{{\textcolor{blue}{\bf \tt  Discrete}}}
\newcommand{\Invariant}{{\textcolor{blue}{\bf \tt  Invariant}}}
\newcommand{\Condition}{{\textcolor{blue}{\bf \tt  Condition}}}
\newcommand{\Init}{{\textcolor{blue}{\bf \tt  Init}}}
\newcommand{\Real}{{\textcolor{blue}{\bf \tt  Real}}}
\newcommand{\real}{{\textcolor{blue}{\bf \tt  real}}}
\newcommand{\Constant}{{\textcolor{blue}{\bf \tt  Constant}}}
\newcommand{\new}{{\textcolor{blue}{\bf \tt  new}}}
\newcommand{\this}{{\textcolor{blue}{\bf \tt  this}}}
\newcommand{\Skip}{{\textcolor{blue}{\bf \tt  Skip}}}

\newcommand{\LRAngle}[1]{\langle #1 \rangle}
\tikzstyle{big_}=[draw,text width=20em,text centered, minimum height=2.5em]
\tikzstyle{big} = [big_, minimum height=3em, rounded corners]

\tikzstyle{auto}=[draw, text width=16em,text centered, minimum height=2.5em]
\tikzstyle{sc} = [auto, minimum height=3em, rectangle split,rectangle split parts=2, rounded corners, draw=green!80!blue!50,very thick,font=\ttfamily,top color=white,bottom color=white]

\tikzstyle{autoc}=[draw, text width=8em,text centered, minimum height=2.5em]
\tikzstyle{scc} = [autoc, minimum height=3em, rectangle split,rectangle split parts=2, rounded corners, draw=green!80!blue!50,very thick,font=\ttfamily,top color=white,bottom color=white]

\tikzstyle{autom}=[draw, text width=13em,text centered, minimum height=2.5em]
\tikzstyle{scm} = [autom, minimum height=3em, rectangle split,rectangle split parts=2, rounded corners, draw=green!80!blue!50,very thick,font=\ttfamily,top color=white,bottom color=white]

\newcommand{\fofo}{\fontsize{6pt}{5pt}\selectfont} 
\newcommand{\fofosmall}{\fontsize{8pt}{5pt}\selectfont}

\makeatletter
\renewcommand\normalsize{%
   \@setfontsize\normalsize\@xpt\@xiipt
   \abovedisplayskip 4\p@ \@plus2\p@ \@minus5\p@
   \abovedisplayshortskip \z@ \@plus3\p@
   \belowdisplayshortskip 6\p@ \@plus3\p@ \@minus3\p@
   \belowdisplayskip \abovedisplayskip
   \let\@listi\@listI}
\makeatother

\newbox\subfigbox % Create a box to hold the subfigure.
\makeatletter
\newenvironment{subfloat}% % Create the new environment.
{\def\caption##1{\gdef\subcapsave{\relax##1}}%
\let\subcapsave=\@empty % Save the subcaption text.
\let\sf@oldlabel=\label
\def\label##1{\xdef\sublabsave{\noexpand\label{##1}}}%
\let\sublabsave\relax % Save the label key.
\setbox\subfigbox\hbox
\bgroup}% % Open the box...
{\egroup % ... close the box and call \subfigure.
\let\label=\sf@oldlabel
\subfigure[\subcapsave]{\box\subfigbox\sublabsave}}%
\makeatother

\begin{document}
\title{Apricot\ --\ An Object-Oriented Modeling Language for Hybrid Systems}

\author{Huixing Fang\inst{1} \and Huibiao Zhu\inst{1} \and Jianqi Shi\inst{2}}
\institute{Shanghai Key Laboratory of Trustworthy Computing\\
Software Engineering Institute, East China Normal University, China\\
\and School of Computing, National University of Singapore, Singapore\\
\email{\{wxfang,hbzhu\}@sei.ecnu.edu.cn},~\email{shijq@comp.nus.edu.sg}}

\maketitle

\begin{abstract}
We propose Apricot as an object-oriented language for modeling hybrid systems.
The language combines the features in domain specific language and object-oriented language, that
fills the gap between design and implementation, as a result, we put forward the modeling language with 
simple and distinct syntax, structure and semantics.
In addition, we introduce the concept of design by convention into Apricot.
As the characteristic of object-oriented and the component architecture in Apricot, 
we conclude that it is competent for modeling hybrid systems without losing scalability.
\keywords Apricot, Object-Oriented Modeling, Hybrid Systems, Design by Convention
\end{abstract}

\section{Introduction}\label{sec:introduction}
\vspace*{-2mm}
Hybrid systems are concerned about the discrete control mode transitions, 
the continuous physical behavior,  and the interaction between these two parts.
As mentioned in \cite{dorf2011modern},
the design of a system is the process that building a concrete to carry out some goals.
Meanwhile, people in the hybrid systems domain have the ambition to control their environment, i.e., 
the physical world. 
For hybrid systems, numerous modeling approaches had been proposed, the hybrid automata \cite{ACHH93,alur1996automatic}, 
 Hybrid CSP \cite{jifeng1994hybrid,ChaoChen96}, HyPA (hybrid process algebra) \cite{cuijpers2005hybrid}, and hybrid program \cite{PlatzerBook}, etc. 
 Regarding the formal verification on hybrid systems, various tools can be used, for instance,
 HyTech \cite{henzinger1997hytech}, d/dt \cite{asarin2002d}, PHAVer \cite{Frehse08}, SpaceEx \cite{FrehseGDCRLRGDM11}, and KeYmaera \cite{platzer2008keymaera}.
These works are respectable and formal,  the common feature is that most of them are focus on the high level  abstraction of hybrid systems.
 However, industrial applications of formal methods need a great level of abstraction 
in existing development processes and an easier manner to adopt for users.
In other words, usability and complexity hiding are the major concerns for 
designers and developers in industry.
Modelica \cite{fritzson1998modelica} is a multi-domain object-oriented modeling language, 
it involves systems relating electrical, mechanical, control, and thermal components, etc. 
And, one of  the characteristics of Modelica is that, the class in Modelica can not be executed explicitly, but simulated by a simulation engine.
From the 1.0 release in 1997 when it began to model continuous dynamic systems to the 3.3
release in May, 2012 the addition of periodic and non-periodic synchronous controllers,
the revision  of Modelica has never been ceased. 
The description capability of Modelica is powerful, and the applications of Modelica is pervasive.
Nevertheless, it is not designed for formal verification, although it is quite  suitable for simulation.
The reason is that the semantics of Modelica is prone to be deterministic, 
however in the area of hybrid systems, it is  prone to consider  the non-deterministic evolution of the system behavior.

\vspace*{-1mm}
The motivation to propose the language Apricot  is that,
we want to construct  an object-oriented  language for modeling hybrid systems.
The language should satisfy the following requirements.
First, clear and simple syntax. 
We know that binary code is accurate and precise, 
so why people in the highly developed modern society do not use 
binary code as the communication language in life. 
Because binary code is closed to hardware, it is far from daily life and hard to be acquainted.
The same is in the area of hybrid systems.
A language that is  close to the designers and developers in industry is needed and worthy to be developed.
Second, distinct structure.  As an object-oriented language, we can employ design patterns \cite{designpatterns1993} in the system design process. For instance, to demonstrate the hierarchical structure of complex hybrid systems, we utilize the composition pattern to build the ownership relation between 
global system and  subsystems. Composition pattern in Apricot constructs the tree structure with respect to objects of {\em System}, {\em Plant}, {\em Dynamic} and the subsystems of {\em Plant} object. We treat objects of {\em Dynamic} and {\em System} as a similar way under the compositional relationship, it results in the ownership between plant
 and subsystem, and then the relationship between system and subsystem.
The third is an explicit semantics.  We propose the  operational semantics for Apricot. As the highly structural style of Apricot models, the semantics is clear and compositional.

The contributions of our work can be elaborated as follows.
The first is about the  innovation on the Interface conception. 
Interface is an abstraction of the type,
  a suitable Interface for hybrid systems should consider the relations for system components and in favor of the 
hierarchical structure construction for complex systems.
The common constraints and conventions are  better to be defined in the abstract level than in the implementation part. Because, the higher the common knowledge is the easier the developer to know well.
 Traditionally, in object-oriented languages, the Interface 
only contains methods and no instance variable declaration or just the constant (in Java, or property in C\#, etc.)  is allowed.  In Apricot, we allow variable requirements, constraint indications and built-in block statements in the Interface. 
The variable requirements define the relationships between the current type and other types. Therefore, it has the ability to describe the ownership among different components.
The constraint indications denotes the behavior that is forced to conform. For instance, the {\em clock} constraint
 indication for the {\em Controller} Interface set the derivative of the variable of {\em Controller} to be the constant number one. 
 The built-in block statement denotes the right usage and position that the block should be.  
In Apricot, for example,  the {\em Condition} block is positioned in the {\em Composition} method of the Interface {\em Plant}.
As a consequence, 
the innovation enhances and clarifies the  relationship for various system components by variable requirements,  specifies the limitation of some components by constraint indications, and explicitly
states the proper usages of blocks by the built-in block statement declaration in Interface. 

Moreover, we apply  the principle of {\em Architecture as Language}, and build the combination of the features from Domain Specific Language (abbreviated as DSL, \cite{DSL2013,fowler2010domain})  and Object-Oriented Language (abbreviated as OOL). The DSL notations (such as the variable requirements and constraint indications) used in Apricot are good for the building of component architecture, and as a result, it makes easier to communicate with domain experts during the system design process. On the other hand, the OOL is familiar to developers in industry, 
and close to the implementations of the system. The combination of DSL and OOL in Apricot 
fill the gap between the design at higher level and the implementation for the concrete.
This paper is organized as follows.
Section \ref{sec:syntax-of-apricot} describes the syntax of Apricot and an example (bouncing ball) 
modeling under Apricot.
The operational semantics is demonstrated in Section \ref{sec:operational-semantics}. 
In addition, Section \ref{sec:design-by-convention} discusses the features of design by convention in Apricot.
And, we make the conclusions in Section \ref{sec:conclusions}.

\begin{figure}
\begin{subfloat}
\begin{minipage}[b]{0.45\textwidth}
\begin{tikzpicture}[rounded corners, thick,shading=ball]
\draw[blue] (0,0) -- (5,0);
\draw[blue] (0,-.05) -- (5,-.05);
\foreach \x in {0,0.07,...,5}
{
\draw (\x,0) -- (\x+.05,-.05);
}
\draw[black, |<->|] (1.5,2.75) -- (1.5,0) node [midway, left, blue] {{\em h}};
\shade[ball color=red!50!green] (2,3) circle (2ex);
\shade[ball color=red!50!green] (3,.28) circle (2ex);
\draw[densely dotted] (3,.28) circle (2ex);
\draw[black,->] (2,2.5) -- (2,2) node [right,blue] { {\em  g}};
\draw[black,->] (3,.28) -- (3,.8) node [right,blue] {{\em R}};
\draw[black,->] (3,.28) -- (3,-.3) node [left,blue] {{\em F}};
\filldraw [black] (3,.28) circle (.5pt);
\end{tikzpicture}
\caption{The ball. {\em h} denotes the height of the ball, {\em g} is for the acceleration of gravity. }
\label{fig:BALLView}
\end{minipage}
\end{subfloat}
~
\begin{subfloat}
\begin{minipage}[b]{0.5\textwidth}
\begin{Verbatim}[commandchars=\\\{\},fontsize=\fofo,numbers=left,numbersep=0pt,frame=lines]
\Dynamic Moving\{
  \Real height,velocity,acceleration;
  /*Constructor*/
  Moving(\Real height,\Real velocity,\Real acceleration)\{
        \this.height=height;
        \this.velocity=velocity;
        \this.acceleration=acceleration;
  \}
  \Continuous()\{
        dot(height,1) == velocity;
        dot(velocity,1) == -acceleration;
   \}
  \Invariant\{
        height in [0,15];
        velocity in [-60,60];
  \};
\}
  \end{Verbatim}
  \end{minipage}
  \caption{Class Moving implements interface Dynamic. }
\label{fig:Dynamic_Moving}
  \end{subfloat}
\begin{subfloat}
\begin{minipage}[b]{0.45\textwidth}
\begin{Verbatim}[commandchars=\\\{\},fontsize=\fofo,numbers=left,numbersep=0pt,frame=lines]
\ParallelAssignment Jump\{
 \Real height,velocity,coefficient;
 /*Constructor*/
 Jump(\Real height,\Real velocity,\Real coefficient)\{
        \this.velocity = velocity;
        \this.height = height;
        \this.coefficient = coefficient;
  \}
  \Discrete()\{
        velocity = -coefficient * velocity;
        height = height;
  \}
\}
\end{Verbatim}
\end{minipage}
\caption{Class Jump implements interface ParallelAssignment.}
\label{fig:ParallelAssignment_Jump}
\end{subfloat}
~
\begin{subfloat}
\begin{minipage}[b]{0.48\textwidth}
\begin{Verbatim}[commandchars=\\\{\},fontsize=\fofo,numbers=left,numbersep=0pt,frame=lines]
\Plant Ball\{
  \Real height,velocity,k,g;
  Ball(\Real height, \Real velocity, \Real k, \Real g)\{
      \this.height = height;
      \this.velocity = velocity;
      \this.k = k;
      \this.g = g;
  \}
  \Dynamic moving = \new Moving(height,velocity,g);
  \Assignment jump = \new Jump(velocity,height,k);
  \Composition()\{
      CompMJ(moving,jump,moving)\{
          \Condition\{  moving.height==0; \};
      \};
  \}
\}
\end{Verbatim}
\end{minipage}
\caption{Class Ball implements interface Plant.}
\label{fig:BouncingBallSystem_Plant}
\end{subfloat}
\begin{subfloat}
\begin{minipage}[b]{0.44\textwidth}
\begin{Verbatim}[commandchars=\\\{\},fontsize=\fofo,numbers=left,numbersep=0pt,frame=lines]
\Controller God\{
  \Real mass,height,velocity,k,t,g;
  God(\Real mass, \Real height, \Real velocity, 
  \Real k, \Real t, \Real g)\{
      \this.mass = mass;
      \this.height = height;
      \this.velocity = velocity;
      \this.k = k;
      \this.t = t;
      \this.g = g;
  \}
  \Dynamic idle = \new \Dynamic()\{
      \Continuous()\{ dot(t,1)==1; \}
  \};
  \Assignment reset = \Skip;

  \Composition()\{
      CompIR(idle,reset,idle)\{
        \Condition\{
         height == 0;
         Resiliency(mass,velocity,k)>mass*g;\};
      \};
  \}
\}
\end{Verbatim}
\end{minipage}
\caption{The controller of bouncing ball system, class God implements interface Controller. }
\label{fig:BouncingBallSystem_Controller}
\end{subfloat}
~
\begin{subfloat}
\begin{minipage}[b]{0.48\textwidth}
  \begin{Verbatim}[commandchars=\\\{\},fontsize=\fofo,numbers=left,numbersep=0pt,frame=lines]
\System BouncingBall\{
 \Real height,velocity,t;
 \Real h[] = \{15, 10, 12\};
 \Real v[] = \{0, 1, 1.5\};
 \Constant \real g=9.8,k=0.6,mass=5;
 \Controller god=\new God(mass,height,velocity,k,t,g);
 \Plant ball = \new Ball(height,velocity,k,g);
 BouncingBall()\{
       god.CompIR || ball.CompMJ;
       god || ball; //maybe not necessary
 \}
 \Init()\{
       height=h[1],velocity=v[1],t=0;
       god.idle.start();
       ball.moving.start();
 \}
\}
\end{Verbatim}
\end{minipage}
\caption{Class BouncingBall implements interface System. }
\label{fig:BouncingBallSystem}
\end{subfloat}
\caption{Bouncing ball model.}
\label{fig:ModelofBBall}
\end{figure}

\section{Syntax of Apricot}\label{sec:syntax-of-apricot}
\vspace*{-2mm}
In this section we will describe the basic syntax of Apricot.
As a  modeling language for hybrid systems, one has to consider the hierarchical
structures of the system to demonstrate the modularity features, 
and also has to propose the definitions of system dynamics with
 the relations between continuous flow and discrete assignments. 
The following recursive definitions have cover the overview of the above ambition.
{\small 
\setlength{\jot}{2.5pt}
\begin{align*}
 System  ::=&  ParaPlants  \parallel  ParaContrs ;
\\
  ParaPlants  ::=&   \parallel_{i=1}^n    Plant_i ; 
\\
 ParaContrs  ::=&  \parallel_{i=1}^m  Controller_i;
\\
 Plant ::=&  AtomicComp  \mid  Comp(  Dynamic ^+, Assignment ^+,  System );
\\
 Controller  ::=&  AtomicComp ;
\\
 AtomicComp  ::=&   Comp(  Dynamic ^+, Assignment ^+) ;
\\
 Assignment  ::=&  SequentialAssignment \mid  ParallelAssignment .
\end{align*}
}
where $n, m \in \mathbb{Z^+}$(positive integers), symbol `$\parallel$' denotes parallel composition.
`$ { Dynamic} ^+$' represents a set of {\em Dynamic} objects, and `$ {Assignment} ^+$'
 has the similar meanings ({\em Assignment} objects).

The system defined here has the point that each system contains one or more plants and controllers. This is different from other approaches or languages such as hybrid automata which do not have this restrict.

\begin{center}
\fbox{\parbox[t]{0.9\textwidth}{
\vspace*{-3mm}
\begin{multicols}{2}
\begin{asparaenum}[(C.1)]
\setlength{\itemsep}{2pt}
\setlength{\itemindent}{2mm}
\setlength{\labelsep}{1mm}
\item \label{syn:continuous}
 \hfill$\fofosmall\begin{aligned}[t]
&Continuous()\{\\
& \hspace{0.5cm} dot(Var_1,Nat_1) == MathExp_1;\\
& \hspace{0.5cm} dot(Var_2,Nat_2)  == MathExp_2;\\
& \hspace{2.5cm} \cdots\\
& \hspace{0.5cm} dot(Var_n,Nat_n)  == MathExp_n;\\
&\}
\end{aligned}$\hfill\null
\item \label{syn:invariant}
\hfill$\fofosmall\begin{aligned}[t]
&Invariant\{\\
& \hspace{0.5cm} Variable_1  ~in~  \lfloor Real_1, Real_1' \rceil;\\
& \hspace{0.5cm} Variable_2  ~in~  \lfloor Real_2, Real_2' \rceil;\\
& \hspace{2cm} \cdots\\
& \hspace{0.5cm} Variable_n  ~in~   \lfloor Real_n, Real_n' \rceil;\\
&\};
\end{aligned}$\hfill\null
\item \label{syn:discrete}
\hfill$\fofosmall\begin{aligned}[t]
&Discrete()\{\\
& \hspace{0.5cm} Variable_1  =  MathExp_1;\\
& \hspace{0.5cm} Variable_2  =  MathExp_2;\\
&  \hspace{1.8cm} \cdots\\
& \hspace{0.5cm} Variable_n  =  MathExp_n;\\
&\}
\end{aligned}$\hfill\null\\
\item \label{syn:condition}
\hfill$\fofosmall\begin{aligned}[t]
&Condition\{\\
&\hspace{0.5cm}    MathExp_1  ~Rel~  MathExp_1';\\
&\hspace{0.5cm}    MathExp_2  ~Rel~  MathExp_2';\\
&\hspace{2.5cm}                \cdots\\
&\hspace{0.5cm}    MathExp_n  ~Rel~  MathExp_n';\\
&\};
\end{aligned}$\hfill\null
\end{asparaenum}
\end{multicols}
}}
\end{center}

{\em Dynamic} object is an instance of the class that implements the  {\em Dynamic}  interface. {\em Dynamic} object is refers to flows which are used to model continuous behavior of physical plants. The implementation class of  {\em Dynamic} interface defines the continuous valuations of the variables in the system over time. And, it also specifies the invariant of the continuous flow. 
The {\em Continuous} method in the {\em Dynamic} implementation class has the form as depicted in (C.\ref{syn:continuous}),
in which, for $1 \leq i \leq n, Var_i$ is the variable of the system, natural number $Nat_i$ represents the derivative order of $Var_i$ that is not equal to 0, $MathExp_i$ is the mathematical expression with the definition:

Let $Vars$ be the set of all variables of system, $\dot{V}ars$ denotes the set of derivative order variables, e.g., if $v \in Vars$, then the first order derivative $\dot{v} \in \dot{V}ars$ ($\dot{v}$ is represented by expression $dot(v,1)$ in Apricot).
{\fofosmall
\begin{align*}
 MathExp ::=  Function(Vars,\dot{V}ars) ;
\end{align*}
}
where, $Function$ defines the mathematical function defined by the designer or the built-in function in Apricot.
Such as addition, subtraction, multiplication, division, etc. For example, the multiplication in Fig.~\ref{fig:ParallelAssignment_Jump} is an infix form function.

The {\em Invariant} statement specifies the properties of the system during the continuous evolution, as illustrated in (C.\ref{syn:invariant}).
In which, $Real$  denotes the real number, $\lfloor \in \{~ '(', '[' ~\}$, and $\rceil \in \{~')', ']'~\}$. Symbols $'(', ~ ')'$ are used to define open intervals, and $'[',~']'$ for closed intervals. For example, in Fig.~\ref{fig:Dynamic_Moving}, `{\tt \small height in [0, 15]}' clarifies the variable {\tt \small height} evaluates the value within the closed interval [0, 15] during the continuous evolution. Note that, the left-open parenthesis is limited to the special real number {\tt \small -Inf}, and the right-open parenthesis is limited to {\tt \small Inf}, thus intervals like 
$(1,2)$,  
$({\tt \small -Inf}, {\tt \small Inf}]$,  
$[{\tt \small -Inf}, {\tt \small Inf})$   and $[{\tt \small -Inf}, {\tt \small Inf}]$ is invalid.

{\em Assignment} interface has two sub-interfaces, {\em SequentialAssignment} and {\em ParallelAssignment}. Both implementations  have a discrete method with the  form in (C.\ref{syn:discrete}).
If this discrete method is defied in class implementing the interface {\em SequentialAssignment}, then it is the sequential composition of these $n$ assignment statements. Otherwise, if it is defined in class implementing the interface {\em ParallelAssignment}, then the parallel composition is the semantics that the assignment statements are supposed to represent. Fig.~\ref{fig:ParallelAssignment_Jump} is an example of {\em ParallelAssignment} implementation.

The {\em Composition} statement  connects the {\em Dynamic} object and {\em Assignment} 
object by a {\em Condition} statement. The {\em Condition} statement has the form in (C.\ref{syn:condition}).
In which, $Rel ~\in \{==,<,>,<=,>=,\text{!=}\}$ is the relation operator, and the expression  ``$MathExp_i~Rel~  MathExp_i';$" defines the relation between the evaluations of $MathExp_i$ and $MathExp_i'$. For example, in Fig.~\ref{fig:BouncingBallSystem_Plant}, the {\em Composition} method refers to
 {\em Dynamic} object {\tt \small moving} and {\em Assignment} object {\tt \small jump}
  with  {\tt \small moving.height==0}. Therefore, if the value of the variable
 {\tt \small height} in {\tt \small moving} is equal to $0$ (i.e., the ball hits the ground), then the {\em Assignment} {\tt \small jump}  will
  be executed and the control will move on to {\tt \small moving}  after this execution
  provided that the invariant is satisfied.

\begin{example}
\label{ex:example_bouncing_ball}
Bouncing ball is a traditional model in hybrid system.  The system has a controller named {\tt \small god} and a plant named {\tt \small  ball}. The controller has {\em Dynamic} {\tt \small idle}, {\em Assignment} {\tt \small reset} and the {\em Composition} relation {\tt \small CompIR} paralleled with plant's {\tt \small CompMJ}. The plant has {\em Dynamic} {\tt \small moving}, {\em Assignment} {\tt \small jump} and the
 {\em Composition} {\tt \small CompMJ} paralleled with controller's {\tt \small CompIR}. The two source-free arrows in the plant
  {\tt \small ball} and controller {\tt \small god} represent the initial dynamics. Therefore, {\tt \small moving} and {\tt \small idle} are the initial dynamics of  {\tt \small ball} and {\tt \small god}, respectively.

  Fig.~\ref{fig:BALLView}--\ref{fig:BouncingBallSystem} are the model code for the bouncing ball system.
  Fig.~\ref{fig:BALLView} depicts the ball, when the ball hits the flat horizontal ground, it suffers the gravity {\em F} and the elastic force {\em R}.
  The class {\tt \small Moving} (in Fig.\ref{fig:Dynamic_Moving}) is an implementation of the {\tt \small Dynamic} interface.
  It declares that the first order derivative of {\tt \small height} over time equals {\tt \small velocity}, 
  and the first order derivative of {\tt \small velocity} over time is equal to {\tt \small -acceleration}. 
  In Fig.~\ref{fig:BouncingBallSystem_Plant}, an object named {\tt \small moving} is created with the type of class {\tt \small Moving}, and relates the variables {\tt \small height}, {\tt \small velocity}, {\tt \small g} of class {\tt \small Ball} to {\tt \small height}, {\tt \small velocity}, {\tt \small acceleration} in class {\tt \small Moving}, respectively.
\end{example}

\subsection{Class, Object and Relation}\label{sec:class,-object-and-relation}

Class declaration defined reference types. The body of class declaration defines
the implementation details.
All classes are non-nested in Apricot. This means that the class declaration defined
within the body of another class or interface is invalid.

The body of a class consists of fields, methods, instance, relations, and constructors.
Field declarations describe instance variables, each instance of the class holds a new 
substantiation of the instance variable. 

\noindent
{\bf Class Declaration.} We have three kinds of class declaration:

\begin{asparaenum}[--]
\setlength{\itemsep}{5pt}
\setlength{\itemindent}{2mm}
\item Top-level Class. If the class do not have super class, and do not implements any other interface:

{\fofosmall
\begin{align*}
& Class ~ Identifier\{ \\
& \hspace{1cm} ClassBody\\
&\}
\end{align*}
}

\vspace*{-4mm}
in which, we do not specify the access modifiers (e.g. {\em Public, Protected, Private} in Java).
The keyword {\tt \small this} in the constructor denotes  the current instance being constructed. If keyword {\tt \small this} occurs in an instance method then it represents the object for which the method was defined. Most of the time, the keyword {\tt \small this} is employed to distinguish the instance variable from parameter variables when the names of variables in different classes clashed.

\vspace*{2mm}
\item { Interface Implementation}.
If one class implements an interface, the class declaration is:

{\fofosmall
\begin{align*}
& InterfaceType ~ Identifier\{ \\
& \hspace{1cm} ClassBody\\
&\}
\end{align*}
}

\vspace*{-4mm}
It is difference from many other object-oriented languages (e.g., Java, C++), we do not use the keyword {\em implements} to  specify the interface type the class implements here.
In example \ref{ex:example_bouncing_ball}, the classes (see Fig.\ref{fig:Dynamic_Moving}--\ref{fig:BouncingBallSystem}) are all interface implementations.

\vspace*{2mm}
\item {Inheritance}.
If one class extends other class (i.e. SuperClass), the class declaration:
{\fofosmall
\begin{align*}
& ClassType ~ Identifier\{ \\
& \hspace{1cm} ClassBody\\
&\}
\end{align*}
}
\end{asparaenum}

\noindent
{\bf Constructor Declaration.}The constructor takes the responsibility for the creation of an instance of a class. Moreover, it weaves the connection between different components in Apricot models. The constructor declaration as follows for the case that formal parameters are presented:

{\fofosmall
\begin{align*}
& Identifier(Formal~Parameters)\{ \\
& \hspace{1cm} ConstructorBody\\
&\}
\end{align*}
}
For example, in Fig.~\ref{fig:BouncingBallSystem_Plant}, the {\tt \small Ball(...)} constructor is:
\vspace*{-2mm}
\begin{Verbatim}[commandchars=\\\{\},fontsize=\fofosmall,frame=single]
            Ball(\Real height, \Real velocity, \Real k, \Real g)\{
                \this.height = height;
                \this.velocity = velocity;
                \this.k = k;
                \this.g = g;
            \}
\end{Verbatim}

The formal parameters are a list of parameter specifiers and separated by the comma symbol `{\tt \small ,}'. Each parameter specifier is a pair of a type and an identifier. The identifier is the name of the parameter.  In Fig.\ref{fig:BouncingBallSystem} line 7, it creates a {\tt \small Ball} object using the `{\tt \small Ball(...)}' constructor. Meanwhile, it creates the connection of variables ({\tt \small height, velocity, k, g}) in {\em system} {\tt \small BouncingBall} with the variables ({\tt \small height, velocity, k, g}) in {\em plant} {\tt \small Ball}. The statements in the constructor of {\tt \small Ball}, e.g. ``{\tt \small this.height = height}'' makes the instance variable {\tt \small height} of {\tt \small Ball} and the instance variable {\tt \small height} of {\tt \small BouncingBall} refer to the same entity. All the modification on variable {\tt \small height} take place in {\tt \small Ball} or {\tt \small BouncingBall} will be recognized immediately by each other.

Formal parameters can be absent, for the case of line 8 in Fig.\ref{fig:BouncingBallSystem}.
The line 9 of the constructor denotes that the composition relation {\tt \small CompIR} of controller {\tt \small god} is parallel with the composition reltion {\tt \small CompMJ} of plant {\tt \small CompMJ}. 
The line 10 denotes that the controller {\tt \small god} is parallel with the plant {\tt \small ball}.
The initializer is declared by the method ``{\tt \small Init}()\{...\}" at line $12 \sim 16$.

\noindent
{\bf Initializer Declaration.} 
The initializer method specifies the initial values of the instance variables in a system. 
For example, the line 13 in Fig.\ref{fig:BouncingBallSystem} sets the initial value of {\tt \small height} to the number 15, {\tt \small velocity} the number 0 and the initial value of 
{\tt \small t} the number 0.
In addition, it  starts the initial dynamics of the components in the system. For instance, 
the initial dynamic of {\em controller} {\tt \small god} is {\tt \small idle} and the initial dynamic of 
{\em plant} {\tt \small ball} is {\tt \small moving} specified by line 14 and line 15 in Fig.\ref{fig:BouncingBallSystem}, respectively.

\noindent
{\bf Anonymous Class Declaration.} 
Anonymous class is an implementation of an interface or an inheritance of a super class.
In Fig.~\ref{fig:BouncingBallSystem_Controller}, the variable {\tt \small idle} declared  at line 12  refers to
an instance of an anonymous class which implements the interface {\tt \small Dynamic}.
The method {\tt \small Continuous}  defined in the anonymous class denotes the first order time-derivative of variable {\tt \small t} is equal to $1$. Therefore, variable {\tt \small t} takes the role 
of a clock. 

Moreover, no invariant is defied in the anonymous class, which means that it has an implicit invariant {\tt \small Ture}, variable {\tt \small t} can take any value in real numbers $\mathcal{R}$.
Anyway, as time is not negative, we can specify an invariant that {\tt \small t} is always equal to or greater than the number 0:
\begin{Verbatim}[commandchars=\\\{\},fontsize=\small]
                       \Invariant\{ t in [0,Inf); \};
\end{Verbatim}
\vspace*{-2mm}
where, `{\tt \small Inf}' denotes the infinity $+\infty$.

\subsection{Interface, Inheritance and Relationship}\label{sec:interface,-inheritance-and-relation} 
In Apricot, there are five built-in interfaces, each defines one key element  of the Apricot model. 
The built-in interface may consist of four parts: method signatures, variable requirements, constraint indications and built-in block statements.  From now on, these four parts are abbreviated to MVCB in this paper.
Method signature defines the name and arguments of the method. 
Variable requirement holds the relations between the current interface and other interfaces, it also restrict the count of objects of the proper types.
Constraint indication demonstrates the limitation for the behavior of the object which implements the interface.
And, the built-in block statement positioned in the interface emphasizes the structure of the language, and indicates 
the right place for the application of the special statement.

\begin{asparaenum}[{\bf --}]
\setlength{\itemsep}{5pt}
\setlength{\itemindent}{2mm}
\item {\bf System Interface} depicted in (I.\ref{itf:system}),
where, `{\em Requires}' is a keyword in Apricot,  `$1..*$' denotes at least one entity. Therefore, each {\em System}  object contains one or more than one {\em Plant} object,  and it  also for the objects of type {\em Controller}.
The method signature `$Init()$' indicates that the {\em System} has an initializer that do not contain any argument and no return value for this initializer.
`{\em plants}' and `{\em controllers}' are the names of the variables referring to the proper types behind the colon symbol (`:').

\item {\bf Plant Interface} depicted in (I.\ref{itf:plant}),
where, it indicates that the implementation of this interface holds several objects of the type {\em Dynamic} and {\em Assignment}, and may have a subsystem or not. 
The {\em Composition} method is used for defining the composition relationships between {\em Dynamic} (or {\em System}) objects and {\em Assignment}  objects.
Each composition relationship with respect to three arguments: the source, action, and the destination.
And, the form `$(dysy[.], ass[.], dysy[.])$' in the composition relationship shows that $dysy[.]$ is the source ({\em Dynamic or System}), ass[.] is the action, and $dysy[.]$ (also can be {\em Dynamic or System}) is the destination, `.' represents the proper index.
The composition relationship denotes the control switch that from the source to the destination under the conditions 
defied in the {\em Condition} block statement. 
During the control switch the action which is restricted to the {\em Assignment} object (i.e., `$ass[.]$') is executed.

\item {\bf Controller Interface} depicted in (I.\ref{itf:controller}),
where, it is the same as {\em Plant} except the {\em Constraint Indication}  and the absent of subsystem.
The {\em clock Constraint Indication}  `$Constraint ~~clock$' denotes that the differential equations in the 
{\em Dynamic}   object of {\em Controller} have the restriction:   the derivative assigned to  the variable is restrict to
  number 1.

\item {\bf Dynamic Interface} depicted in (I.\ref{itf:dynamic}),
where, it indicates that each {\em Dynamic} implementation has a method and an  built-in {\em Invariant} block statement.
The method `$Continuous()$' with respect to the continuous evolution of the system states. The form of the method 
has been declared before in Sect \ref{sec:syntax-of-apricot}. The {\em Invariant} is applied to define the range of proper 
variable concerned for the current {\em Dynamic} object.

\item {\bf Assignment Interface} depicted in (I.\ref{itf:assignment}),
the {\em Assignment} interface only has the method `$Discrete()$'. The {\em Discrete} method
plays the role of the actions that would be executed during the control switch of dynamics.
Moreover, there are two interfaces inherit the {\em Assignment} interface, {\em SequentialAssignment} 
and {\em ParallelAssignment}. {\em SequentialAssignment} has the semantics of sequential composition  for its assignment statements, and  {\em ParallelAssignment} has a parallel composition semantics.
\end{asparaenum}

\vspace*{-3mm}
\begin{center}
\fbox{\parbox[t]{.98\textwidth}{
\vspace*{-3mm}
\begin{multicols}{2}
\begin{asparaenum}[({I}.1)]
\setlength{\itemsep}{3pt}
\setlength{\itemindent}{-1mm}
\setlength{\labelsep}{0mm}
\item \label{itf:system}
\hfill$\fofosmall\begin{aligned}[t]
& Interface ~~ System\{ \\
& \hspace{0cm} Requires ~~ plants[1..*]: Plant;\\
& \hspace{0cm} Requires ~~ controllers[1..*]: Controller;\\
& \hspace{0cm} Init();\\
& \}
\end{aligned}$\hfill\null
\item \label{itf:plant}
\hfill$\fofosmall\begin{aligned}[t]
& Interface ~~ Plant\{ \\
& \hspace{0cm} Requires ~~ dy[1..*]: Dynamic;\\
& \hspace{0cm} Requires ~~ ass[1..*]: Assignment;\\
& \hspace{0cm} Requires ~~sy[0..1]: System;\\
& \hspace{0cm} Composition()\{ \\
& \hspace{0.1cm} Requires ~ coms[1..*] : (dysy[.],ass[.],dysy[.])\\
&         \hspace{1cm}  \{ ~  Condition\{\}   ; ~\};\\
& \hspace{.1cm} \};\\
&\}
\end{aligned}$\hfill\null
\item \label{itf:controller}
\hfill$\fofosmall\begin{aligned}[t]
& Interface ~~ Controller\{ \\
& \hspace{0cm} Constraint ~~ clock;\\
& \hspace{0cm} Requires ~~ dy[1..*]: Dynamic;\\
& \hspace{0cm} Requires ~~ ass[1..*]: Assignment;\\
& \hspace{0cm} Composition()\{ \\
& \hspace{0.1cm} Requires ~ coms[1..*] : (dy[.],ass[.],dy[.])\\
&         \hspace{1cm}  \{~  Condition\{\};\};\\
& \hspace{0.1cm} \};\}
\end{aligned}$\hfill\null
\item \label{itf:dynamic}
$\fofosmall\begin{aligned}[t]
& Interface ~ Dynamic\{ \\
& \hspace{0cm} Continuous();\\
& \hspace{0cm} Invariant\{\};\\
& \}
\end{aligned}$\hfill\null
\item  \label{itf:assignment}
$\fofosmall\begin{aligned}[t]
& Interface ~ Assignment\{ \\
& \hspace{0cm} Discrete();\\
& \}
\end{aligned}$\hfill\null
\end{asparaenum}
\end{multicols}
}}
\end{center}

In addition, as the existence of MVCB in the interface declaration, we claim that the inheritance of class or interface in Apricot should consider to inherit and follow the MVCB in the super-class or super-interface. And, the implementation of interface in Apricot should consider to implement and follow the MVCB in the implemented interface.

\section{Operational Semantics}\label{sec:operational-semantics}
\vspace*{-2mm}
Structural operational semantics (\cite{plotkin1981structural,Plotkin04a}, SOS) was proposed by G.D.Plotkin in 1981.  Transition system is the base for structural operational semantics. It takes
the transition relation between configurations to characterize the operational feature of  system behaviour.  Usually, SOS is applied to the programs and operations on discrete data. In order to deal with continuous data, we need to abstract the continuous features, and then obtain a discrete view of the continuous data for hybrid system. For the semantics and verification of object-oriented languages, some related works can be found in \cite{america1986operational,apt2011verification,jifeng2006rcos}.

\begin{definition}
A Transition System ({\bf TS}) is a structure consists of a set of configurations ({\bf C}) and the relation ($\rightarrow$) between configurations, i.e., $\bf TS \overset{\text{def}}{=} \langle C, \rightarrow \rangle$, where $\bf \rightarrow \subseteq C \times C$.
\end{definition}

\vspace*{-8mm}
\subsection{Configurations}\label{sec:configurations}
\vspace*{-2mm}
Any insight into a hybrid system is obtained through the state of the system.
 Each state is a valuation of the variables in the system. After the system start-up, it always accompanied with a state at each time point. All the states compose a state space of the system. Based on the state space, one can
check whether some specific state can be reached by the system for some proper initial states. It is called the reachability analysis. And, various respectable works had been done, e.g., the Hytech \cite{henzinger1997hytech} proposed by Henzinger etc., the Phaver \cite{Frehse08} and SpaceEx \cite{FrehseGDCRLRGDM11} by Frehse etc., the hybrid process algebra approach \cite{cuijpers2005hybrid} by P.J.L. Cuijpers, and Platzer's dynamic differential logic \cite{PlatzerBook}, etc.

Besides system states, to reveal the relation between statement and state,
 we also need to pay attention to the statements (control flow) throughout the system execution. 
These understanding can be used to check the statement-related properties. 
For example, we can check that some particular dynamic method is not reached or executed by the system with the knowledge of both statement and state.

\begin{definition}
We define the set of configurations with statements, states, and types, formally as follows:
{\fofosmall
\begin{align*}
{\bf C}::=& \LRAngle{\mathcal{P}(\Theta), \mathcal{P}(\Sigma), \mathcal{P}({\bf T})},\\
\Theta ::=& \{\vartheta_1.\vartheta_2.\cdots.\vartheta_n \mid \vartheta_{i} \text{ is a statement of Apricot}\},\\
\Sigma ::=& {\bf Vars} \times {\bf Vals},\\
\bf T ::=& {\bf Vars} \times {\bf Types},
\end{align*}}
where $1\leq i \leq n$, $\Theta$ denotes the set of prefix annotated statements, $\mathcal{P}(\Theta)$ is the power set of $\Theta$, $\Sigma$ is consists of all functions that mapping from the set of variables {\bf Vars} to the set of values {\bf Vals}, {\bf T} is a set of functions which relate each variable in {\bf Vars} with a type in {\bf Types}. 
\end{definition}

\begin{figure}
\centering
\begin{tikzpicture}[scale=0.7]
\tikzstyle{every node}=[scale=0.8]
\path 
node at (0,3.5) (system) [scm] {System \nodepart[color=blue]{two} {\em \color{red} system}};
\path
node at (0,0.6) (init) [scm] {init() \nodepart[color=blue]{two} {\em {\bf system}.\color{red}init()}};
\path
node at (7.5,3.5) (assign) [scm] {height=h[1] \nodepart[color=blue]{two} {\em {\bf system.init()}.\color{red}height=h[1]}};
\path
node at (7.5,1) (start) [sc] 
{
god.idle.start() \\
ball.moving.start()
\nodepart[color=blue]{two} 
{\em {\bf system.init()}.{\color{red}god.idle.start()}\\
{\color{blue} {\bf system.init()}}.\color{red}ball.moving.start()}
};
\draw[rounded corners=2mm,|->,line width=1pt,red!50!black!50] (system) -- (init.north);
%\path (system) edge[thick,->,bend right=30] (init.west);
\draw[rounded corners=2mm,|->,line width=1pt,red!50!black!50]   (init.east)  -- +(1,0.4) |- (assign.west);
%\path (init.east) edge[thick,->,bend right] (assign);
\draw[rounded corners=2mm,|->,line width=1pt,red!50!black!50]  (assign.south) -- (start);
%\path (assign.east) edge[thick,->,bend left] (start);
\end{tikzpicture}
\vspace*{-3mm}
\caption{The example of prefix annotated statements for bouncing ball system. The italic statement  is the current statement the system executed. }
\label{fig:prefixannotatedstatement}
\end{figure}
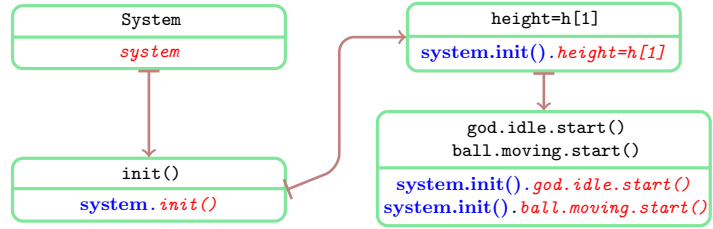

A prefix annotated statement is a linked list
that begins with a variable ($\vartheta_1$) which denotes the system and ended with the statement ($\vartheta_n$) currently executed or expression to be evaluated. Along the list there
will be objects or methods.  An Apricot model comprises more than one component, and  these components paralleled. As a result, the first element of a configuration is a subset of $\Theta$,  consists of the parallel prefix annotated statements. (Fig. \ref{fig:prefixannotatedstatement} illustrates the example prefix annotated statements for bouncing ball system)

\vspace*{-5mm}
\vspace*{-5mm}

Moreover, considering the nondeterminism feature of Apricot, a model of Apricot consists of numerous prefix annotated statements, thus all the possible runs of the model can be
illustrated by a tree structure, and each branch may has a different state space. 

%Fig.~\ref{fig:prefixannotatedstatement} is an overview of the prefix annotated statements for the bouncing ball system in example \ref{ex:example_bouncing_ball}, provided that {\tt \small system} is refers to the system object.

%\begin{figure}[tbph]
%\centering
%\includegraphics[width=0.5\linewidth]{./prefixannotatedstatement}
%\caption{The  tree structure of prefix annotated statements for bouncing ball system.}
%\label{fig:prefixannotatedstatement}
%\end{figure}

\subsection{Axioms and Rules}\label{sec:axioms}
\vspace*{-2mm}
Here, we will give the axioms for Apricot. 
Consider single statement $\theta$, for $\{{\bf Pre}.\theta\} \in \mathcal{P}(\Theta)$, $\sigma \in \mathcal{P}(\Sigma)$, and $\tau \in \mathcal{P}(\bf T)$, then $\LRAngle{\{{\bf Pre}.\theta\}, \sigma, \tau} \in {\bf C}$. For simplicity, we take 
 $\bf Pre.\theta$ for $\{{\bf Pre}.\theta\}$  in the following axioms  ({\bf Pre} is the prefix):
 
\begin{asparaenum}[{\bf --}]
\setlength{\itemsep}{5pt}
\setlength{\itemindent}{2mm}
\item {\bf Arithmetic expression $e$}.

    Evaluation of constant numbers:
    
    {\fofosmall
    \begin{align}
    \LRAngle{{\bf Pre}.n,\sigma,\tau} \rightarrow n, 
    \end{align}
    }
\text{where $n$ is a constant number.}

Evaluation of variable:

{\fofosmall
      \begin{align}
        \LRAngle{{\bf Pre}.v,\sigma,\tau} \rightarrow n, 
        \end{align}
}        
\text{where, $v$ is a variable of number type, and $\sigma(v)=n$.}
    
Evaluation of addition:
    
{\fofosmall
    \begin{align}
    \frac{\LRAngle{{\bf Pre}.e_1,\sigma,\tau} \rightarrow n_1 ~~ \LRAngle{{\bf Pre}.e_2,\sigma,\tau} \rightarrow n_2}{\LRAngle{{\bf Pre}.(e_1+e_2),\sigma,\tau} \rightarrow n}
    , 
    \end{align}
}   
where, $e_1$ and $e_2$ are variables or constant numbers, and  $n$ is the summation of $n_1$ and $n_2$.

\item {\bf Mathematical function expression}.

Derivative over time $t$ with order $n$:

{\fofosmall
\begin{align}
\label{derivativeovertime}
    \LRAngle{{\bf Pre}.dot(v,n),\sigma,\tau} \rightarrow \frac{d^n v}{dt^n}, 
\end{align}
}
where, $\frac{d^n v}{dt^n}$ is a formula that represents the $n$-th order derivative of $v$ 
over time. In fact, we can regard the $n$-th order derivative as an attribute or observation of the variable, and 
employ a new variable to maintain the value of the derivative. We produce a new variable when it occurs at the 
first time, and the name would be $v\_n$. Thus, (\ref{derivativeovertime}) is changed to

{\fofosmall
\begin{align}
\frac{\LRAngle{v\_n,*} \notin \sigma}{\LRAngle{{\bf Pre}.dot(v,n),\sigma,\tau} \rightarrow \LRAngle{{\bf Pre}.v\_n,\sigma',\tau'}}
    , 
\end{align}
}
where,  symbol `*'  stands for any value, $\sigma'=\sigma[v\_n:=null]$, $\tau'=\tau[v\_n:=\tau(v)]$. And, if $v_n$ is already in $\sigma$, then
we have:
{\fofosmall
\begin{align}
\frac{\LRAngle{v\_n,*} \in \sigma}{\LRAngle{{\bf Pre}.dot(v,n),\sigma,\tau} \rightarrow v\_n}
    , 
\end{align}
}
Derivative over other variable $u$ with order $n$:
{\fofosmall
\begin{align}
    \LRAngle{{\bf Pre}.dot(v,u,n),\sigma,\tau} \rightarrow \frac{d^n v}{du^n}, 
\end{align}
}
and, if $v\_y\_n$ is new, then we have
{\fofosmall
\begin{align}
\frac{\LRAngle{v\_y\_n,*} \notin \sigma}{\LRAngle{{\bf Pre}.dot(v,y,n),\sigma,\tau} \rightarrow \LRAngle{{\bf Pre}.v\_y\_n,\sigma',\tau'}}
    , 
\end{align}
}
otherwise,
{\fofosmall
\begin{align}
\frac{\LRAngle{v\_y\_n,*} \in \sigma}{\LRAngle{{\bf Pre}.dot(v,y,n),\sigma,\tau} \rightarrow v\_y\_n}
    .
\end{align}
}
\item {\bf Assignment}.

For single assignment,
 {\fofosmall
    \begin{align}
    \LRAngle{{\bf Pre}.(v=e),\sigma,\tau} \rightarrow \LRAngle{{\bf Pre}.skip,\sigma',\tau} , 
    \end{align}
}
where, $v$ is a variable, $e$ is for arithmetic expression, and the updated state $\sigma'=\sigma[v:=\sigma(e)]$.

For sequential assignment and parallel assignment, 
consider the assignment statements in the {\em Discrete} method $S$:

{\fofosmall
\begin{align*}
	 	Discrete()\{
	 	     x = y;
	 	     y = x;
	 	\}
\end{align*}
}
\begin{asparaenum}
\setlength{\itemindent}{5mm}
\item As Sequential Assignment:  executing $S$ in a state with $x=0$ and $y=1$, $x$ and $y$ are both evaluate to the value  $1$. For assignment statements $S_1, S_2$ in Sequential Assignment method,

 {\fofosmall
    \begin{align}
\frac{\LRAngle{{\bf Pre}.S_1,\sigma,\tau} \rightarrow \LRAngle{{\bf Pre}.skip,\sigma',\tau}}{
    \LRAngle{{\bf Pre}.(S_1;S_2),\sigma,\tau} \rightarrow \LRAngle{{\bf Pre}.S_2,\sigma',\tau}} , 
    \end{align}
}    
\item As Parallel Assignment:  executing $S$ in the same state, $x$ and $y$ exchange their value, $x$ is changed to $1$, $y$ is $0$. For assignment statements $S_1, S_2$ in Parallel Assignment method, $v_1$ is the variable modified by $S_1$ and $v_2$ of $S_2$,

 {\fofosmall
    \begin{align}
    \frac{
    \begin{array}{c}
\LRAngle{{\bf Pre}.S_1,\sigma,\tau} \rightarrow \LRAngle{{\bf Pre}.skip,\sigma',\tau},
\LRAngle{{\bf Pre}.S_2,\sigma,\tau} \rightarrow \LRAngle{{\bf Pre}.skip,\sigma'',\tau} \\
    \end{array}
    }{\LRAngle{{\bf Pre}.(S_1||S_2),\sigma,\tau} \rightarrow \LRAngle{{\bf Pre}.skip,\sigma''',\tau} , }
    \end{align}
}    
    where, $\sigma'''=\sigma[v_1:=\sigma'(v_1), v_2:=\sigma''(v_2)]$, `$||$' denotes that the assignments ($S_1,S_2$) in {\em Discrete} 
    method of {\em ParallelAssignment} object are executed in parallel.
\end{asparaenum}

\item {\bf Method Invocation}.

\begin{asparaenum}
\setlength{\itemindent}{5mm}
\item Zero-Arity-Argument method $m()$:

{\fofosmall
    \begin{align}
    \LRAngle{{\bf Pre}.m(),\sigma,\tau} \rightarrow \LRAngle{{\bf Pre}'.S,\sigma,\tau}, 
    \end{align}
}    
where, ${\bf Pre}'={\bf Pre}.m$ and $S$ is the body of method $m$.

\item Fixed-Arity-Argument method $m(arg[1..n])$:

 {\fofosmall
    \begin{align}
    \LRAngle{{\bf Pre}.m(exp[1..n]),\sigma,\tau} \rightarrow \LRAngle{{\bf Pre}'.S,\sigma',\tau'}, 
    \end{align}
}    
where, ${\bf Pre}'={\bf Pre}.m(exp[1..n])$ and $S$ is the body of method $m$,  
for $1 \leq i \leq n$, $arg[i]$ is a new variable, and,
{\fofosmall
\begin{align*}
\sigma'=&\sigma[ arg[i]:=\sigma(exp[i]) ],
\end{align*}
}
if $\tau(exp[i])$ is a subtype of the defined type of $arg[i]$, then
{\fofosmall
\begin{align*}
\tau'=&\tau[ arg[i]:=\tau(exp[i]) ], 
\end{align*}
}
otherwise, $\tau'(arg[i])$ takes the defined type of the formal parameter.
\end{asparaenum}

\item {\bf Instance variable}. Suppose $var$ is an instance variable of the object $obj$.
\begin{asparaenum}
\setlength{\itemindent}{5mm}
\item Declaration of instance variable without initialization. Consider the declaration $D$:
{\fofosmall
\begin{align*}
Type~~var;
\end{align*}
}
This  defines a variable $var$ of type $Type$ and assigns the special value $null$ to $var$.
Thus, we have
 {\fofosmall
    \begin{align}
    \LRAngle{{\bf Pre}.obj.D,\sigma,\tau} \rightarrow  \LRAngle{{\bf Pre}.obj.Skip,\sigma',\tau'}, 
    \end{align}
}    
where, $\sigma'=\sigma[var:=null]$ and $\tau'=\tau[var:=Type]$.

\item Declaration of instance variable with initialization. Consider the declaration $D$:
{\fofosmall
\begin{align*}
Type~~var = val;
\end{align*}
}
This  defines a variable $var$ of type $Type$ and assigns the  value $val$ to $var$.
Thus, we have
{\fofosmall
    \begin{align}
    \LRAngle{{\bf Pre}.obj.D,\sigma,\tau} \rightarrow  \LRAngle{{\bf Pre}.obj.Skip,\sigma',\tau'}, 
    \end{align}
}
where, $\sigma'=\sigma[var:=val]$ and 
if $\tau(val)$ is a subtype of $Type$, then $\tau'=\tau[var:=\tau(val))]$, otherwise, 
$\tau'=\tau[var:=Type]$.
\end{asparaenum}

\item {\bf Local variable}. Suppose $var$ is a local variable in the method $m$ or block $b$.
The following are demonstrated under the scenario with method $m$.
\begin{asparaenum}
\setlength{\itemindent}{5mm}
\item Declaration of local variable without initialization. Consider the declaration $D$:
{\fofosmall
\begin{align*}
Type~~var;
\end{align*}
}
This  defines a variable $var$ of type $Type$ and assigns the special value $null$ to $var$.
Thus, we have
 {\fofosmall
    \begin{align}
    \LRAngle{{\bf Pre}.m.D,\sigma,\tau} \rightarrow  \LRAngle{{\bf Pre}.m.Skip,\sigma',\tau'}, 
    \end{align}
 }
where, $\sigma'=\sigma[var:=null]$ and $\tau'=\tau[var:=Type]$.

\item Declaration of local variable with initialization. Consider the declaration $D$:
{\fofosmall
\begin{align*}
Type~~var = val;
\end{align*}
}
This  defines a variable $var$ of type $Type$ and assigns the  value $val$ to $var$.
Thus, we have
 {\fofosmall
    \begin{align}
    \LRAngle{{\bf Pre}.m.D,\sigma,\tau} \rightarrow  \LRAngle{{\bf Pre}.m.Skip,\sigma',\tau'}, 
    \end{align}
}    
where, $\sigma'=\sigma[var:=val]$ and 
if $\tau(val)$ is a subtype of $Type$, then $\tau'=\tau[var:=\tau(val))]$, otherwise, 
$\tau'=\tau[var:=Type]$.

\item End of method $m$. 
{\fofosmall
    \begin{align}
    \LRAngle{{\bf Pre}.m.End,\sigma,\tau} \rightarrow  \LRAngle{{\bf Pre}.Skip,\sigma',\tau'}, 
    \end{align}
}    
 where, $\sigma'=\sigma[return:=\sigma(returnExp), rm~vars]$ and $\tau'=\tau[return:=\tau(returnExp), rm~vars]$. `$rm~vars$' represents 
 the removing of all the mappings related to local variables of the method $m$. 
 $End$ denotes the end of the method, usually a method is ended by 
 explicitly a {\em Return} statement or the right brace `\}' positioned at the end of the method body. $return$ is the special variable refers to the result of the method invocation,
 $returnExp$ denotes the value of the variable. And, for block $b$, the special variable $return$ is ignored.
\end{asparaenum}

\item {\bf Object Creation}. The procedure of object creation is composed of instance variable initialization and constructor invocation.

\begin{asparaenum}
\setlength{\itemindent}{5mm}
\item  Creation by Constructor. If the object creation statement $S$ is
{\fofosmall
\begin{align*}
Type~obj = new~M(exp[0..n]);
\end{align*}
}
where, $exp[0..n]$ represents the list (or array) of actual parameters. $M$ 
is the name of the instantiated class, also the name of the constructor,
 $M(exp[0..n])$ is an invocation of the corresponding constructor in 
the class. Suppose the set of instance variable declaration is $Ds$, and constructor $m(arg[0..n])$,
{\fofosmall
    \begin{align} \label{objectcreation}
    \LRAngle{{\bf Pre}.S,\sigma,\tau} \rightarrow  \LRAngle{{\bf Pre}'.(Ds;M(exp[0..n])),\sigma',\tau'}, 
    \end{align}
}
where, ${\bf Pre}'={\bf Pre}.obj$, $\sigma'=\sigma[obj:=o]$, $o$ is a new object of $Type$ with all the instance variables refer to the special value $null$, $\tau'=\tau[obj:=Type]$.

\item  Creation by Anonymous Class.  If the object creation statement $S$ is
{\fofosmall
\begin{align*}
Type~obj = new~Identifier()\{Class~Body\};
\end{align*}
}
Then, 
{\fofosmall 
    \begin{align}
    \LRAngle{{\bf Pre}.S,\sigma,\tau} \rightarrow  \LRAngle{{\bf Pre}'.(Ds;M()),\sigma',\tau'}, 
    \end{align}
}
where, it is the same as (\ref{objectcreation}) except that it executes the zero-arity-argument constructor.
If there is no  zero-arity-argument constructor declares in the class body, the empty one would take the job that doing nothing when  it is invoked. The empty constructor is implicitly declared in one class for  the case that the  zero-arity-argument constructor is missing by the designer.
\end{asparaenum}

\item {\bf Dynamic}. 
In Apricot, the {\em Dynamic} object consists of one {\em Continuous} method and an {\em Invariant} block.
The {\em Continuous} method declares the differential equations that the dynamic flow followed with respect to the properties defined within the {\em Invariant} block. 
The properties in the {\em Invariant} block indicate the  range of the variables during the continuous evolution.
For dynamic, if the dynamic flow reaches the border of the {\em Invariant} and all the conditions of the
compositions from the dynamic can not be satisfied,  
then the control is waiting at the border provided that  any advancement  of 
the flow according to the {\em Continuous} method will violate the {\em Invariant}.

\begin{asparaenum}
\setlength{\itemindent}{5mm}
\item Differential Equation.
For  one statement $D$ that is declared in the $Continuous$ method, $D$ is a differential equation for the variable $v$.  
For variable $v$ and nature number $n$, mathematic expression $me$, the differential equation $D$ is 

{\fofosmall
\begin{align*}
dot(v,n) == me;
\end{align*}
}
Suppose that there exists a function $f: I \to \mathbb{R}$, and $I$ is a time-interval $[a, b]$, i.e.,  the domain of $f$,
and  the value of $v$ at time-point $t \in [a,b]$ is $f(t)$.
Here,  the start time-point of the continuous evolution following $D$ is at time $a$, the end point  $b$ is for some proper time-point greater than or equals $a$.
Then, before the termination of the flow, at some time-point $t \in [a, b]$, we have

 {\fofosmall
    \begin{align}
    \LRAngle{{\bf Pre}.D,\sigma,\tau} \rightarrow  \LRAngle{{\bf Pre}.D,\sigma',\tau}, 
    \end{align}
}    
    where, $\sigma'=\sigma[v:=f(t)]$. We call $f$ the {\em Real-Function} for $D$, and $t$ the {\em Proper-Time}.
    
 \item Termination of Flow.  The dynamic flow reaches the border of the {\em Invariant} and no valid composition relationship exists,  
 then the control is waiting at the border if the forward flow would violate the {\em Invariant}.
 
 {\fofosmall
     \begin{align}
     \frac{
     \begin{array}{c}
 \forall c \in C,\LRAngle{{\bf Pre}.c,\sigma,\tau} \rightarrow False,\\
 \LRAngle{{\bf Pre}.D,\sigma,\tau} \rightarrow  \LRAngle{{\bf Pre}.D,\sigma',\tau},
 \exists i \in I, \LRAngle{{\bf Pre}.i,\sigma',\tau} \rightarrow False \\
     \end{array}
     }{\LRAngle{{\bf Pre}.D,\sigma,\tau} \rightarrow  \LRAngle{{\bf Pre}.(dot(tw,1)=1),\sigma,\tau} },
     \end{align}
     }
 where, the set $C$ is the $Condition$ block related to the current $Dynamic$ object that contains the  differential equation $D$. And, $I$ is the $Invariant$ block in the $Dynamic$ object, it is the set of conditions should be satisfied during the continuous evolution.  For $\forall t \in (a,b]$, $\sigma'=\sigma[v:=f(t)]$, in which, $f: I \to \mathbb{R}$, $I = [a, b]$, $f(t)$ is the value of $v$ at the time-point $t \in I$.
At last, $tw$ is a  specific variable for  the waiting time after the flow terminated.

\end{asparaenum}
\item {\bf Invariant}. 
An Invariant block $I$ is a built-in block in a  {\em Dynamic} object.
Actually, $I$ consists of  conditions. Each condition specifies the
 range of one variable, and can be evaluated as a Boolean expression.
 Suppose $i \in I$, and $i \equiv v ~in~ (e_1, e_2)$, we have
 {\fofosmall
\begin{align}
\frac{
\begin{array}{c}
\LRAngle{{\bf Pre}.(e_1,e_2),\sigma,\tau} \rightarrow (n_1,n_2),
\sigma(v) \in (n_1, n_2)\\
\end{array}
}
{\LRAngle{{\bf Pre}.(v ~in~ (e_1, e_2)),\sigma,\tau} \rightarrow True}
   , 
\end{align}
}
where, $v$ takes the value in the interval denoted by $(e_1, e_2)$. 
And, the opposite situation,
{\fofosmall
\begin{align}
\frac{
\begin{array}{c}
\LRAngle{{\bf Pre}.(e_1,e_2),\sigma,\tau} \rightarrow (n_1,n_2),
\sigma(v) \notin (n_1, n_2)\\
\end{array}
}
{\LRAngle{{\bf Pre}.(v~in~(e_1, e_2)),\sigma,\tau} \rightarrow False}
   .
\end{align}
}
Now, we have the evaluation of an Invariant $I$ based on the up two laws,
{\fofosmall
\begin{align}
\frac{
\begin{array}{c}
\forall  i \in I, \LRAngle{{\bf Pre}.i,\sigma,\tau} \rightarrow True
\end{array}
}
{\LRAngle{{\bf Pre}.I,\sigma,\tau} \rightarrow True}
   , 
\end{align}
}
where, $I$ is true when all the conditions in it is true. And, if there exists an invalid condition, then
$I$ is false,
{\fofosmall
\begin{align}
\frac{
\begin{array}{c}
\exists  i \in I, \LRAngle{{\bf Pre}.i,\sigma,\tau} \rightarrow False
\end{array}
}
{\LRAngle{{\bf Pre}.I,\sigma,\tau} \rightarrow False}
   .
\end{align}
}
\item {\bf Condition}.  A {\em Condition} block $C$ consists of a number of Boolean expressions.
Each Boolean expression $c$ involves two mathematic expressions ($me_1, me_2$) and a relational operator $opt$.
Let $c \equiv me_1 ~opt~ me_2$, $opt \in \{==, <, >, <=, >=, !=\}$, and $C$ for the set of all Boolean expressions 
in the {\em Condition} block,  
{\fofosmall
\begin{align}
\frac{
\begin{array}{c}
\forall  c \in C, \LRAngle{{\bf Pre}.c,\sigma,\tau} \rightarrow True
\end{array}
}
{\LRAngle{{\bf Pre}.C,\sigma,\tau} \rightarrow True}
   , 
\end{align}
}
where, $C$ is true iff all Boolean expressions in $C$ is true.
\item {\bf Composition Relationship}. 
It involves the control switch from one dynamic to another under proper conditions.
Let $D_1, D_2$  represent two {\em Dynamic} objects, they may be the same object, e.g., in
Example \ref{ex:example_bouncing_ball}.
And, let $C$ be one of the {\em Condition} blocks related to $D_1$ and $D_2$.
For {\em Composition Relationship} $CR$, and the corresponding {\em Assignment} object $A$, let
$R$ be the name of the {\em Composition Relationship}, then

{\fofosmall
\begin{align*}
CR \equiv R(D_1,A,D_2)\{C\}.
\end{align*}
}
For convenience, we simplify it to
{\fofosmall
\begin{align*}
CR \equiv R(D_1,A,D_2, C).
\end{align*}
}
Thus, we have the valid composition relationship, 

{\fofosmall
     \begin{align}
     \frac{
     \begin{array}{c}
 \LRAngle{{\bf Pre}.C,\sigma,\tau} \rightarrow True,
 \LRAngle{{\bf Pre}.A,\sigma,\tau} \rightarrow  \LRAngle{{\bf Pre}.Skip,\sigma',\tau},
  \LRAngle{{\bf Pre}.D_2.I,\sigma',\tau} \rightarrow True \\
     \end{array}
     }{\LRAngle{{\bf Pre}. R(D_1,A,D_2, C),\sigma,\tau} \rightarrow True},
     \end{align}
}     
where, $I$ is the {\em Invariant} of $D_2$. 
And, the control switch from $D_1$ to $D_2$ may occurs when the relationship is valid,
{\fofosmall
\begin{align}
\frac{\LRAngle{{\bf Pre}. R(D_1,A,D_2, C),\sigma,\tau} \rightarrow True}
{\LRAngle{{\bf Pre}. D_1,\sigma,\tau} \rightarrow \LRAngle{{\bf Pre}. D_2,\sigma',\tau}}.
\end{align}
}
Note that, the control switch may not take place even though the relationship is valid.
It means that, if the {\em Invariant} of  $D_1$ is true and $D_1$ can continue the continuous evolution without to violate the {\em Invariant}, then the choice to switch or continue the flow itself is 
nondeterministic.

\item {\bf Start Dynamics}.
For {\em Dynamics} $D_1$ and $D_2$, the composite for start statements, is the parallel evolution of 
the continuous flows, let
{\fofosmall
\begin{align*}
D_1||D_2 \equiv D_1.start(); D_2.start(),
\end{align*}
}
then, we have
 {\fofosmall
     \begin{align}
     \frac{
     \begin{array}{c}
 			\LRAngle{{\bf Pre}.D_1,\sigma,\tau} \rightarrow \LRAngle{{\bf Pre}.D_1,\sigma_1,\tau},
 			\LRAngle{{\bf Pre}.D_2,\sigma,\tau} \rightarrow \LRAngle{{\bf Pre}.D_2,\sigma_2,\tau}  \\
     \end{array}
     }{\LRAngle{{\bf Pre}.(D_1||D_2),\sigma,\tau} \rightarrow \LRAngle{{\bf Pre}.(D_1||D_2),\sigma',\tau}},
     \end{align}
  }
where, $\sigma_1=\sigma[v_1:=f_1(t)]$, and $\sigma_2=\sigma[v_2:=f_2(t)]$, therefore,
{\fofosmall
\begin{align*}
\sigma'=  \sigma_1[v_2:=f_2(t)] = \sigma_2[v_1:=f_1(t)] =  \sigma[v_1:=f_1(t), v_2:=f_2(t)].
\end{align*}
}
Here, $f_1, f_2$ are the {\em Real-Function}s for $D_1$ and  $D_2$, respectively. 
And, $t$ is the {\em Proper-Time}.

\item {\bf Parallel Composition Relationship}. For two composition relationships $CR_s$ and $CR_t$,
the parallel composition relationship is defined as follows,
{\fofosmall
\begin{align*}
CR_s ~||~ CR_t \equiv R_s(D_{s_1},A_s,D_{s_2},C_s) ~||~  R_t(D_{t_1},A_t,D_{t_2}, C_t).
\end{align*}
}
First, we have the parallel execution of {\em Assignment} objects $A_s$ and $A_t$,
{\fofosmall
\begin{align}
\frac{
\begin{array}{c}
\LRAngle{{\bf Pre}.D_{s_1},\sigma,\tau} \rightarrow \LRAngle{{\bf Pre}.D_{s_2},\sigma',\tau},
\LRAngle{{\bf Pre}.D_{t_1},\sigma',\tau} \rightarrow \LRAngle{{\bf Pre}.D_{t_2},\sigma'',\tau}
\end{array}
}
{
\begin{array}{c}
\LRAngle{{\bf Pre}.(A_s || A_t),\sigma,\tau} \rightarrow \LRAngle{{\bf Pre}. Skip,\sigma'',\tau} \\
\end{array}
},
\end{align}
}
where, $A_s$ and $A_t$ are symmetrical. The parallel composition relationship is valid, 
{\fofosmall
 \begin{align}
     \frac{ 
     \begin{array}{c}
 			\LRAngle{{\bf Pre}.(CR_s ~and~ CR_t),\sigma,\tau} \rightarrow True,\\
 			\LRAngle{{\bf Pre}.(A_s || A_t),\sigma,\tau} \rightarrow \LRAngle{{\bf Pre}. Skip,\sigma'',\tau},
            \LRAngle{{\bf Pre}.(D_{s_2}.I ~and~ D_{t_2}.I),\sigma'',\tau} \rightarrow True,
     \end{array}
     }
     {\LRAngle{{\bf Pre}.(CR_s || CR_t),\sigma,\tau} \rightarrow True},
\end{align}
}
where, the Boolean operator $and$ represents the conjunction relation.

\end{asparaenum}

\section{Design by Convention}\label{sec:design-by-convention}
\vspace*{-2mm}
Design by convention is a software design paradigm that is known as convention over configuration (abbreviated as COC).
It evicts the decisions the developers need to make by the conventional usages of the design ingredients, given the simplicity during the modeling process.
In software development, COC is usually used for the least configuration that 
the developer should to set down. We apply the idea of COC and utilize it in the design of hybrid systems, and name it as design by convention (abbreviated as DBC) in our language.

\subsection{The composition of statements}
\vspace*{-2mm}
For boolean expressions $A$ and $B$, 
{\small
\begin{align*}
Condition\{A;B;\}; \equiv A \wedge B.
\end{align*}}
We do not need to explicitly add the conjunction  operation to connect the boolean expressions, the separate expressions in the {\em Condition} block have the conjunction relationship implicitly. It also makes the conditions more
clear and be easy to  understand.

For the parallel and sequential assignments, they have the same appearance, but, different execution semantics indicated by the different Interfaces.
{\small
\begin{align*}
ParallelAssignment\{Discrete()\{A;B;\}\}; \equiv A || B,\\
SequentialAssignment\{Discrete()\{A;B;\}\}; \equiv A ; B.
\end{align*}}
The implementation of Interface {\em ParallelAssignment} gives the statements $A$ and $B$ the parallel composition relationship. While, the sequential composition of $A$ and $B$ is prominent for the case of Interface {\em SequentialAssignment}.

In a similar way, the starts of dynamics in the {\em Initializer} method for the {\em System} class have the parallel composition semantics without to employ the parallel operator `$||$'.
{\fofosmall
\begin{align*}
Init\{A.start();B.start();\}; \equiv A || B
\end{align*}}
And, in the constructor of a {\em System} class, we can ignore the parallel indications for plants and controllers if 
they have the starts of dynamics in the {\em Initializer}. For instance, the `{\tt \small god||ball}' in Fig.\ref{fig:BouncingBallSystem} can be wiped off.
\subsection{The inexistence}
\vspace*{-2mm}
For True Condition and Invariant,
{\small
\begin{align*}
Condition\{\}; \equiv True, Invariant\{\}; \equiv True.
\end{align*}}
We evaluate the empty {\em Condition} and {\em Invariant} blocks to {\em True}, and the inexistent of these two blocks also considered to the boolean {\em True}.

For {\em Empty} assignment or the non-initialization of the assignment instance variable, we evaluate it to the special statement {\em Skip}.
{\small
\begin{align*}
Comp(Dy_1, , Dy_2) \equiv Comp(Dy_1, Skip , Dy_2),
\end{align*}}
where $Dy_1$ and $Dy_2$ are dynamics and the ` ' (Blank Space) in the LHS denotes the empty assignment.

\section{Conclusions}\label{sec:conclusions}
\vspace*{-2mm}
In this paper, we proposed Apricot as an object-oriented language for modeling hybrid systems and described the syntax and operational semantics of Apricot in detail.
The language combines the features from DSL and OOL, that
fills the gap between design and implementation, as a result, bring about a modeling language with 
simple and distinct syntax, structure and semantics.
We also discussed the design by convention features of Apricot.
 For the future work, we will focus on the formal verification 
for Apricot models,  then  investigate verification techniques and develop relevant tools.

\bibliographystyle{splncs03}
\bibliography{bibApricot}

\begin{thebibliography}{10}
\providecommand{\url}[1]{\texttt{#1}}
\providecommand{\urlprefix}{URL }

\bibitem{ACHH93}
Alur, R., Courcoubetis, C., Henzinger, T., Ho, P.: Hybrid automata: An
  algorithmic approach to the specification and analysis of hybrid systems. In:
  Hybrid Systems, LNCS, vol. 736, pp. 209--229. Springer-Verlag (1993)

\bibitem{alur1996automatic}
Alur, R., Henzinger, T., Ho, P.: Automatic symbolic verification of embedded
  systems. IEEE Transactions on Software Engineering  22(3),  181--201 (1996)

\bibitem{america1986operational}
America, P., de~Bakker, J., Kok, J., Rutten, J.: Operational semantics of a
  parallel object-oriented language. In: Proceedings of POPL'86. pp. 194--208.
  ACM (1986)

\bibitem{apt2011verification}
Apt, K., De~Boer, F., Olderog, E., de~Gouw, S.: Verification of object-oriented
  programs: a transformational approach. Journal of Computer and System
  Sciences  (2011)

\bibitem{asarin2002d}
Asarin, E., Dang, T., Maler, O.: The d/dt tool for verification of hybrid
  systems. In: Proceedings of CAV'02, LNCS, vol. 2404, pp. 365--370.
  Springer-Verlag (2002)

\bibitem{cuijpers2005hybrid}
Cuijpers, P.J.L., Reniers, M.A.: Hybrid process algebra. The Journal of Logic
  and Algebraic Programming  62(2),  191--245 (2005)

\bibitem{dorf2011modern}
Dorf, R.C., Bishop, R.H.: Modern Control Systems. Prentice Hall (2011)

\bibitem{fowler2010domain}
Fowler, M.: Domain-specific languages. Addison-Wesley Professional (2010)

\bibitem{Frehse08}
Frehse, G.: Phaver: algorithmic verification of hybrid systems past hytech.
  International Journal on Software Tools for Technology Transfer  10(3),
  263--279 (2008)

\bibitem{FrehseGDCRLRGDM11}
Frehse, G., Guernic, C.L., Donz{\'e}, A., Cotton, S., Ray, R., Lebeltel, O.,
  Ripado, R., Girard, A., Dang, T., Maler, O.: {SpaceEx: Scalable verification
  of hybrid systems}. In: Proceedings of CAV'11. LNCS, vol. 6806, pp. 379--395.
  Springer-Verlag (2011)

\bibitem{fritzson1998modelica}
Fritzson, P., Engelson, V.: Modelica -- a unified object-oriented language for
  system modeling and simulation. In: Proceedings of ECOOP'98, LNCS, vol. 1445,
  pp. 67--90. Springer-Verlag (1998)

\bibitem{designpatterns1993}
Gamma, E., Helm, R., Johnson, R.E., Vlissides, J.M.: Design patterns:
  Abstraction and reuse of object-oriented design. In: Proceedings of ECOOP'93.
  LNCS, vol. 707, pp. 406--431. Springer-Verlag (1993)

\bibitem{jifeng1994hybrid}
He, J.: From csp to hybrid systems, a classical mind: essays in honour of car
  hoare (1994)

\bibitem{jifeng2006rcos}
He, J., Li, X., Liu, Z.: rcos: A refinement calculus of object systems.
  Theoretical Computer Science  365(1),  109--142 (2006)

\bibitem{henzinger1997hytech}
Henzinger, T., Ho, P., Wong-Toi, H.: Hytech: A model checker for hybrid
  systems. International Journal on Software Tools for Technology Transfer
  1(1),  110--122 (1997)

\bibitem{PlatzerBook}
Platzer, A.: Logical Analysis of Hybrid Systems - Proving Theorems for Complex
  Dynamics. Springer (2010)

\bibitem{platzer2008keymaera}
Platzer, A., Quesel, J.D.: Keymaera: A hybrid theorem prover for hybrid systems
  (system description). In: Automated Reasoning, pp. 171--178. Springer-Verlag
  (2008)

\bibitem{plotkin1981structural}
Plotkin, G.D.: A structural approach to operational semantics. Tech. Rep. DAIMI
  FN-19, Department of Computer Science, Aarhus university (September 1981)

\bibitem{Plotkin04a}
Plotkin, G.D.: A structural approach to operational semantics. The Journal of
  Logic and Algebraic Programming  60-61,  17--139 (2004)

\bibitem{DSL2013}
Voelter, M., Benz, S., Dietrich, C., Engelmann, B., Helander, M., Kats, L.C.L.,
  Visser, E., Wachsmuth, G.: DSL Engineering - Designing, Implementing and
  Using Domain-Specific Languages. dslbook.org (2013)

\bibitem{ChaoChen96}
Zhou, C., Wang, J., Ravn., A.P.: A formal description of hybrid systems. In:
  Hybrid Systems III, LNCS, vol. 1066, pp. 511--530. Springer-Verlag (1996)

\end{thebibliography}

\appendix

\section{Identifiers}
\label{sec:identifiers}

An {\em identifier} is an unlimited-length (but the length is greater than one) sequence of letters and digits, but not a Keyword:
\begin{eqnarray*}
 Letter  &::=& {\tt a \mid b \mid ... \mid z \mid A \mid B \mid ... \mid Z} ;\\
 Digit  &::=& {\tt 1 \mid 2 \mid 3 \mid4 \mid5 \mid6 \mid7 \mid8 \mid9 \mid0};\\
 ValidChar  &::=&  Letter   ~\mid~  Digit ;\\
 Identifier  &::=&   Letter  \{ Letter  ~ \mid~  Digit  \}^*.
\end{eqnarray*}
In which the letter is defied as the character in the set 
$\{${\tt a}, {\tt b}, {\tt c}, {\tt d}, 
{\tt e}, {\tt f}, {\tt g}, {\tt h}, 
{\tt i}, {\tt j}, {\tt k}, {\tt l}, {\tt m}, 
{\tt n}, {\tt o}, {\tt p}, {\tt q}, {\tt r}, 
{\tt s}, {\tt t}, {\tt u}, {\tt v}, {\tt w}, 
{\tt x}, {\tt y}, {\tt z}, {\tt A}, {\tt B}, 
{\tt C}, {\tt D}, {\tt E}, {\tt F}, {\tt G}, 
{\tt H}, {\tt I}, {\tt J}, {\tt K}, {\tt L}, 
{\tt M}, {\tt N}, {\tt O}, {\tt P}, {\tt Q}, 
{\tt R}, {\tt S}, {\tt T}, {\tt U}, {\tt V}, 
{\tt W}, {\tt X}, {\tt Y}, {\tt Z}
$\}$.

\section{Types, Values, and Variables}\label{sec:types,-values,-and-variables}
The types of the Apricot  language are divided into two categories:
mathematic types and reference types. 
\begin{align*}
 Type ::= & PrimitiveType \mid MathematicType  \\
          & \mid  ReferenceType ;\\
\end{align*}

\subsection{Mathematic Types and Values}
Primitive Type is the same as mathematicType except that, primitive type variable 
can not be shared and has the feature of ``call-by-value" during method calls. 
Call-by-value requires the evaluation of the arguments before passing them to the definition of the method. Another style is call-by-name which passing the arguments directly to the definition.
For mathematic and reference types we take the call-by-name style argument passing for method invocation. In addition, there is a difference between mathematic type and reference type. Reference type variables can refer to another object with the same type by the assignment statement. But, the assignment can only change the mathematical value of the object for mathematic type variables. It means that, when a mathematic type variable refers to a methematic type object for the first time, the variable will hold this object all the time and only the mathematical value of this object can be updated.
\begin{eqnarray*}
 MathematicType  &::=&  NumbericType  \mid {\tt Boolean};\\
 NumbericType  &::=& {\tt Integer} \mid {\tt Real};
\end{eqnarray*}
Accordingly, the primitive type is defined by:
\begin{eqnarray*}
 PrimitiveType &::=&  {\tt integer} \mid {\tt real} \mid {\tt boolean} ;\\
\end{eqnarray*}

\begin{enumerate}
\item Mathematic types : {\tt Boolean} type and the numeric types. The {\tt Boolean}
type represents a logical quantity in the literals set \{ {\tt True}, {\tt False}\}.
The numeric types are the integer type {\tt Integer}, and the real number type
{\tt Real};
\item Reference types : class types, interface types, and array types.
\end{enumerate}

An object is a dynamically created instance of a class type or a dynamically 
created array. The values of a reference type are references to objects.

\subsection{Reference Types and Values}

There are four kinds of reference types: class types, interface types, type variabless,
and array types.
\begin{align*}
 ReferenceType  ::=&  ClassType  \mid  InterfaceType   \\
                   &  \mid  ArrayType ;\\
 ClassType  ::=&  Identifier ;\\
 InterfaceType  ::=&  Identifier  \mid {\tt System} \mid {\tt Plant} \\
& \mid  {\tt Controller}\\
& \mid {\tt Dynamic} \mid {\tt Assignment} \\
& \mid {\tt ParallelAssignment} \\
& \mid {\tt SequentialAssignment} \\
 ArrayType  ::=&  Type  ~[ ~ ].
\end{align*}

\subsection{Variables}

A variable is a physical quantity name in physical world or a storage location in the memory of computer, and has an associated type that is either a mathematic type or a reference type.

The value of a variable  is changed by an assignment or according to the differential equations defined in {\tt Dynamic} classes.

For all types, the default value of any type variable is the special value {\tt null}.

\subsection{Variables of Mathematic Type}
Mathematic type variables are always hold a mathematic value of that exact mathematic type.

\subsection{Variables of Reference Type}
A variable of a reference type {\tt R} can hold a null reference, a reference to an instance of class {\tt C}, any class that is a subclass of {\tt C}, any class that is a implementation of interface {\tt C} or any array type.

\section{Mathematical Operations}\label{sec:mathematical-operations}

\subsection{Arithmetic Operators}
For $x,y \in \mathcal{R}$, the following arithmetic operators are defined on Real numbers ($\mathcal{R}$):
\begin{enumerate}
\item  $x + y$,  binary plus, addition;
\item  $x - y$,  binary minus,subtraction;
\item  $x * y$,  binary multiple, multiplication;
\item  $x / y$,  binary divide, division;
\item  $+x$, unary plus, it denotes the identity operation on $x$, thus, $x == +x$ with
 respect to the evaluation;
\item  $-x$,  unary minus, inverse operation on $x$, thus, $(-x) + x == 0$.
\end{enumerate}

\subsection{Boolean Operators}
Standard boolean operators are defined for all {\tt Boolean} type values $x, y$:
\begin{enumerate}
\item $==$, equality;

\item $!=$,  inequality;

\item $!$, logical complement;

\item {\tt in},  belong to interval, the result value of ($x$ {\tt in} $(a,b)$) is {\tt True} iff $a < x < b$,
 ($x$ {\tt in} $[a,b]$) is {\tt True} iff $a \leq x \leq b$,
($x$ {\tt in} $(a,b]$) is {\tt True} iff $a < x \leq b$, and 
 ($x$ {\tt in} $[a,b)$) is {\tt True} iff $a \leq x < b$;

\item {\tt and}, the result value of ($x$ {\tt and} $y$) is {\tt True} if both operand values are {\tt True};

\item {\tt xor}, the result value of ($x$ {\tt xor} $y$) is {\tt True} if the operand 
values are different;

\item {\tt or}, the result value of ($x$ {\tt or} $y$) is {\tt True} if one of the operand values is {\tt True}.
\end{enumerate}

\subsection{Numeric Comparisons}
 Standard comparison operations are defined for all Real numbers ($\mathcal{R}$), which result in a value of type {\tt Boolean}:
\begin{enumerate}
\item $==$,   equality;
\item $!=$,   inequality;
\item $<$,   less than;
\item $<=$,   less than or equal to;
\item $>$,   greater than;
\item $>=$,   greater than or equal to.

\end{enumerate}

Special Symbol numbers:
\begin{enumerate}
\item $Inf$ is stands for $\infty$, which is equal to itself and greater than any other number;
\item $-Inf$ is stands for $-\infty$, which is equal to itself and less then any other number;
\end{enumerate}

\subsection{Mathematical Functions}
We provides a comprehensive collection of mathematical functions and operators. These mathematical operations are defined on Real numbers ($\mathcal{R}$).
\begin{enumerate}
\item $dot(x,n)$, n-th order derivative of $x$ over time ($t$), i.e. $dot(x,n)=\frac{d^n x}{dt^n}$.
\item $dot(x,y,n)$, n-th order derivative of $x$ over $y$, i.e. $dot(x,y,n)=\frac{d^n x}{d y^n}$.
\item Standard trigonometric functions: $sin$,    $cos$,    $tan$,    $cot$,    $sec$ and    $csc$.
\item $ round(x) $, round $x$ to the nearest integer, omitting decimal fractions smaller than $0.5$, e.g. $round(2.5)=3$, $round(0.4)=0$.
\item $ floor(x) $, round $x$ towards $-Inf$, e.g. $round(2.5)=2$.
\item $ ceil(x) $, round $x$ towards $+Inf$, e.g. $ceil(2.5)=3$.
\item $ div(x,y) $, truncated division, and quotient rounded towards zero.
\item $ fld(x,y) $, floored division, quotient rounded towards $-Inf$.
\item $ rem(x,y) $, remainder, satisfies $x = div(x,y)*y + rem(x,y)$, implying that sign of $rem(x,y)$ matches $x$.
\item $ mod(x,y) $, modulus; satisfies $x = fld(x,y)*y + mod(x,y)$, implying that sign of $ mod(x,y)$ matches $y$.
\item $ gcd(x_1,x_2,...,x_n) $, greatest common divisor of $x_1$, $x_2$, ..., $x_n$ with sign matching $x_1$.
\item $ lcm(x_1,x_2,...,x_n) $, least common multiple of $x_1$, $x_2$, ..., $x_n$ with sign matching $x_1$.
\item $ abs(x) $, a positive value with the magnitude of $x$.
\item $ sign(x) $, indicates the sign of $x$, returning $-1$, $0$, or $+1$.
\item $ sqrt(x) $, the square root of $x$, i.e. $x^2$.
\item $ root(x,b) $, the b-th root of $x$, i.e. $\sqrt[b]{x}$.
\item $ hypot(x,y) $, accurate $sqrt(x^2 + y^2)$ for all values of $x$ and $y$.
\item $ pow(x,y) $, $x$ raised to the exponent $y$, i.e. $x^y$.
\item $ exp(x) $, the natural exponential function at $x$, i.e. $e^x$.
\item $ log(x) $, the natural logarithm of $x$, i.e. $\log(x)$ or $\ln(x)$.
\item $ log(b,x) $, the base b logarithm of $x$, i.e. $\log_b(x)$.
\item $ erf(x) $, the error function (Gauss error function) at $x$, i.e. $erf(x)=\frac{2}{\sqrt{\pi}}\int_0^x{e^{t^2}dt}$.
\item $ gamma(x) $, the gamma function at $x$.
\item $max(x_1,...,x_n)$.
\item $min(x_1,...,x_n)$.
\end{enumerate}

\end{document}